\documentclass[aps,prb,twocolumn,showpacs,floats,preprintnumbers,amsmath,amssymb]{revtex4}

\usepackage{graphicx}
\usepackage{dcolumn}
\usepackage{bm}
\usepackage{subfigure}

\begin{document}

\preprint{(to be submitted to Phys. Rev. B)}

\title{Assessing the efficiency of first-principles basin-hopping sampling}

\author{Ralf Gehrke}
\author{Karsten Reuter}

\affiliation{Fritz-Haber-Institut der MPG, Faradayweg 4-6, D-14195 Berlin (Germany)}

\date{\today}

\begin{abstract}
We present a systematic performance analysis of first-principles basin-hopping (BH) runs, with the target to identify all low-energy isomers of small Si and Cu clusters described within density-functional theory. As representative and widely employed move classes we focus on single-particle and collective moves, in which one or all atoms in the cluster at once are displaced in a random direction by some prescribed move distance, respectively. The analysis provides detailed insights into the bottlenecks and governing factors for the sampling efficiency, as well as simple rules-of-thumb for near-optimum move settings, that are intriguingly independent of the distinctly different chemistry of Si and Cu. At corresponding settings, the observed performance of the BH algorithm employing two simple, general-purpose move classes is already very good, and for the small systems studied essentially limited by frequent revisits to a few dominant isomers.
\end{abstract}

\pacs{Valid PACS appear here}

\maketitle

\section{Introduction}

Research on small particles containing up to a few tens of atoms is largely driven by their novel properties that are significantly affected by (quantum) size effects, particularly in the interplay between structural and electronic degrees of freedom.\cite{clusters} Such clusters, thus, carry the potential of major technological advances for applications exploiting their already exemplified unique optical, magnetic, and chemical properties. Atomically resolved structural information is a key prerequisite towards employing these envisioned functionalities, considering that the latter will be tailored to the atomic scale. In this respect not only the ground state isomer will be of importance, but potentially all energetically low-lying metastable isomers.

A materials modeling targeting the identification of such relevant cluster isomers involves the global and local exploration of the corresponding vast configuration space, suitably represented by the high-dimensional potential-energy surface (PES) \cite{wales03} $E(\{{\bf R}_m\})$ where ${\bf R}_m$ is the position of atom $m$ in the cluster. The rapid growth of the number of local PES minima, i.e. metastable isomers, with increasing cluster size quickly limits approaches focusing only on structural motifs provided by chemical intuition. Required are instead more systematic unbiased sampling techniques and, among those (see e.g. Refs. \onlinecite{kirkpatrick83,szu87,deaven95,deaven96,wolf98,goedecker04}), approaches based on the basin-hopping (BH) \cite{li87,wales97,doye98,wales99,wales00} idea are widespread. In this idea the configuration space is explored by performing consecutive jumps from one local PES minimum to another. To achieve this, positions of atom(s) in the cluster are randomly perturbed in a so-called trial move, followed by a local geometry optimization which brings the system again into a local PES minimum.  

\begin{figure}
\centering
\includegraphics[width=3.8cm,angle=-90]{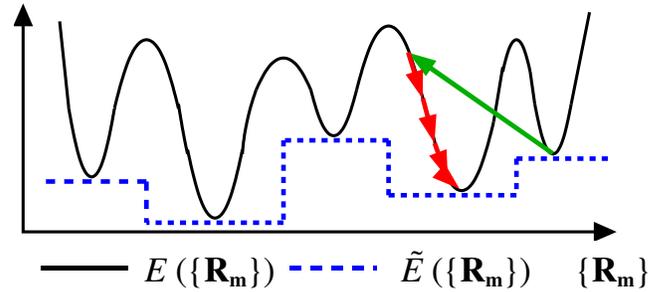}
\caption{(Color online) Schematic representation of the original and transformed potential energy surface, $E(\{{\bf R}_m\})$ and $\tilde E(\{{\bf R}_m\})$ respectively, as well as of a basin-hopping trial move (see text).}
\label{fig1}
\end{figure}

Rather than exploring $E(\{{\bf R}_m\})$, BH approaches concentrate therefore on the transformed PES $\tilde E(\{{\bf R}_m\})$, where the energy at any point in configuration space is assigned to that of the local minimum obtained by the given geometry optimization technique. This maps the PES onto a set of interpenetrating staircases with plateaus, or basins of attraction, corresponding to the set of configurations which lead to a given minimum after optimization. As apparent from Fig. \ref{fig1} the resulting PES topography significantly facilitates interbasin transitions, which constitutes already part of the reason for the success and efficiency of the BH method. In its classical form, BH employs a Metropolis criterion based on an effective temperature $T_{\rm eff}$ to either accept or reject the jump into the PES minimum reached by the trial move. This generates a canonic ensemble on $\tilde E(\{{\bf R}_m\})$, and introduces therewith both the desired importance sampling of the energetically lowest-lying isomers and the possibility to surmount barriers on multiple-funnel type PESs.\cite{doye98} Obvious ramifications of this basic acceptance rule are e.g. to either further promote the downhill driving force to the global minimum by applying a simulated annealing type sequential reduction of $T_{\rm eff}$ during the run, or to extend the importance sampling to all isomers in an energy window above the ground-state by unconditionally accepting all trial isomers with energies in a range above the lowest-energy isomer identified at any given moment in the run. 

When envisioning a predictive and material-specific modeling the accuracy of the PES underlying the sampling is of central importance. Due to the already mentioned intricate coupling of structural and electronic degrees of freedom in small clusters, the nature of the PES must be quantum-mechanic. Compared to simple analytic model potentials corresponding first-principles electronic structure calculations come at a high computational cost, even when describing electronic exchange and correlation only on the level of density-functional theory (DFT) with semi-local functionals. This dictates utmost efficiency of the employed sampling to reduce the number of required energy and force evaluations to the absolute minimum. Apart from the acceptance criterion, the efficiency of the BH method is predominantly governed by the recipe with which trial moves are generated. Among the phletora of move types suggested in the literature many contain technical parameters that are unspecified and which one would correspondingly seek to optimize to reduce the computational cost of a first-principles sampling run. Moreover, rather than revealing inefficient settings only {\em a posteriori}, this optimization would best be carried out by monitoring on-the-fly analyzable performance indicators that allow to adapt an ongoing run. 

Unfortunately, there are few to none general prescriptions of how to set technical move parameters that do not require detailed system-specific insight. With respect to on-the-fly performance indicators there exists at best the rule-of-thumb to aim at an overall acceptance of new trial structures of roughly one half \cite{wales97,frenkel02}. However, this rule emerges from the empirical observation that a factor one half ensures an efficient sampling of canonic ensemble averages and thus must not necessarily carry over to the intended goal of searching for the energetically lowest lying isomers with the least possible number of energy and force evaluations. A second complication arises from the stochastic nature of the BH algorithm. Any analysis measuring the efficiency of technical BH settings or the reliability of suggested on-the-fly performance indicators therefore necessarily needs to involve an averaging over a sufficiently large number of different BH runs starting from different initial structures and using different random number seeds. This would not be too much of a problem when using numerically undemanding model potentials, but then it would be unclear whether the obtained findings are meaningful for proper quantum-mechanic PESs. A straightforward evaluation based on first-principles energetics, on the other hand, is hitherto computationally involved even when only considering smaller clusters up to say 10 atoms.

In this situation, the aim of the present study is to establish a corresponding framework for a systematic performance analysis of first-principles BH sampling runs. An important ingredient herein is the use of a hopping matrix type concept that provides not only a valuable analysis tool, but also helps to bring down the computational cost for the manifold of first-principles BH runs required in the averaging procedure. Using DFT within the generalized gradient approximation to describe the PES, we illustrate the scheme for Si clusters as a system with more directional, covalent type of bonding and for Cu clusters as representative of a metallic system. As a typical example of move classes involving technical parameters we concentrate on so-called single-particle and collective moves, in which either a single randomly chosen atom or all atoms in the cluster at once are displaced in a random direction by some prescribed move distance, respectively. For small clusters up to 10 atoms, our analysis indicates that these moves still enable efficient jumps anywhere in configuration space, i.e. between any PES minima, so that the actual BH acceptance criterion becomes less important. The thereby disentangled influence of move class and acceptance criterion allows us to separately assess the algorithm performance solely with respect to the technical move parameters, here the move distances. The analysis of the obtained results clearly identifies the governing factors and bottlenecks for the sampling efficiency of the investigated small systems, and gives indications on how they scale with increasing cluster size. Apart from providing detailed insights for the specific move classes studied, this stimulates ideas with respect to on-the-fly adaptive settings and establishes a protocol to benchmark more specialized move types.

\section{Theory}

\subsection{Density-Functional Theory}

The underlying PESs are obtained from DFT calculations within the generalized gradient approximation \cite{perdew96} as implemented in the all-electron full-potential code FHI-aims \cite{aims}. In order to suppress a potential complication in the performance analysis due to the spin degrees of freedom all calculations were consistently carried out in a non-spin polarized way. In FHI-aims the Kohn-Sham orbitals are expanded in basis sets consisting of numeric atom-centered orbitals. All calculations reported here were conducted with the so-called ``minimal+$spd$'' basis set. For each considered system we recomputed all stable cluster isomers within an energy range up to 1\,eV above the ground-state, namely those listed in Figs. \ref{fig4}-\ref{fig6} below, also with hierarchically constructed larger basis sets available in FHI-aims. From these calculations we deduce that the relative energies between these isomers are converged to within 10\,meV at the ``minimal+$spd$'' basis set level, which is fully sufficient for the arguments and conclusions put forward below. We also ran several test BH runs with larger basis sets, but never obtained isomers other than those already revealed at the ``minimal+$spd$'' level. This suggests that not only the local minima, but also the other parts of the PES are sufficiently described with the employed ``minimal+$spd$'' basis.

Local structural optimization is done using the Broyden-Fletcher-Goldfarb-Shanno method \cite{recipes}, relaxing all force components to smaller than $10^{-2}$\,eV/{\AA}. While this tight force criterion typically ensures structural convergence to below $10^{-3}$\,{\AA}, it is virtually impossible to converge the DFT total energies up to the number of digits required to uniquely distinguish different isomers from each other. We therefore use the difference norm of all interatomic distances in the cluster as additional tool for the comparison of isomer structures. Two isomers $A$ and $B$ are considered to be equivalent if  
\begin{equation}
  \frac{ \sum_i \left( d_{A,\{i\}} -  d_{B,\{i\}} \right)^2 }{\sum_i \left(
  d_{A,\{i\}}^2 + d_{B,\{i\}}^2 \right)} < \Delta ,
\end{equation}
where $d_{A,\{i\}}$ and $d_{B,\{i\}}$ are the sorted interatomic distances of the two isomers to compare. The denominator serves as normalization which yields a dimensionless quantity that is furthermore species- and cluster-size independent. $\Delta$ can be tuned such that all isomers in the energy range of interest are unambigously distinguished and was taken as $10^{-4}$. In order to check whether the thus identified different isomers are true local minima and not saddle points, they 
were subjected after the BH run to a vibrational analysis based upon a Hessian matrix obtained by finite differences of the analytical atomic forces when displacing all atoms by $10^{-3}$\,{\AA}.

\subsection{Basin-Hopping}

The BH runs explore the configuration space through a sequence of jumps from one PES minimum to another. For this, an initially random cluster structure (created in the spirit of the big-bang method \cite{leary97,yang06}) is subject to so-called trial moves, which correspond to a random structural modification, followed by a local relaxation as depicted in
Fig. \ref{fig1}. As representative and widely used move classes we focus in this work on single-particle and collective moves, in which either a single randomly chosen atom or all atoms in the cluster are randomly displaced, respectively. The corresponding displacement vector of atom $m$ is suitably expressed in spherical coordinates as
\begin{equation}
  \Delta {\bf R}_m = r_m {\bf e}(\theta, \phi) \quad ,
\end{equation}
where ${\bf e}(\theta, \phi)$ is a unit vector in the displacement direction defined by the angles $\theta$ and $\phi$ with respect to an arbitrary, but fixed axis. For an unbiased sampling $\theta \in [0,\pi ]$ and $\phi \in [ 0, 2\pi]$ must be obtained as uniformly distributed random numbers. On the contrary, the move distance $r_m$ is {\em a priori} not specified, but will sensitively determine the jumps in configuration space and therewith the algorithmic performance. It provides therefore a nice example of a technical parameter that one would like to optimize for a first-principles sampling run, yet without introducing bias or system-specific insight. It is furthermore {\em a priori} not clear whether it is preferable to focus on one optimum move distance or whether it possibly advantageous for the overall sampling to include partly shorter and partly longer moves. We study this by drawing the move distances as random numbers distributed around some average value $r_{\rm o}$ $a$, where $a$ is the computed dimer bond length and $r_{\rm o}$ 
correspondingly a less system-dependent unitless quantity. A preference for one optimum distance can then be evaluated by considering a peaked distribution centered around $r_{\rm o}$, whereas the effect of a wide variation of move distances can be tested with a distribution that allows for a broader range of 
values. Specifically, we use either a normal distribution (width $0.07\sqrt{r_{\rm o}}$) around $r_{\rm o}$ for the prior and a uniform distribution (width $r_{\rm o}$) centered around $r_{\rm o}$ for the latter. The goal is therefore to assess the dependence of the sampling efficiency on $r_{\rm o}$ and the form of the distribution around it. In all of these cases an additional important factor is to prevent an entropy-driven dissociation of the cluster during the BH run. We achieve this by disregarding trial moves as well as local relaxations that generate loosely connected or partly dissociated structures characterized by an atom having a nearest-neighbor distance larger than twice the dimer bond length. Similarly discarded are moves that place atoms at distances of less than 0.5\,{\AA} from each other.

Apart from the move class the second fundamental ingredient that needs to be specified in a BH run is the acceptance criterion according to which a generated trial structure is accepted and replaces the current cluster structure as starting point for the following trial move. In order to introduce a downhill driving force towards the energetically low-lying (and ultimately ground-state) isomers it is clear that a more stable trial structure should be unconditionally accepted. In its classical form, the BH scheme also accepts less stable trial structures according to a Metropolis rule, $~ {\rm exp}(- \Delta \tilde{E} / k_{\rm B} T_{\rm eff})$, where $k_{\rm B}$ is the Boltzmann constant, $\Delta \tilde{E} > 0$ the energy difference to the new trial structure, and introducing another unspecified technical parameter which may crucially affect the algorithmic performance, the effective temperature $T_{\rm eff}$. The original motivation behind this Metropolis rule is that the finite possibility to climb uphill enables the algorithm to effectively surmount high-energy barrier regions on multiple-funnel type transformed PESs $\tilde E(\{{\bf R}_m\})$. However, as long as the employed move class enables efficient jumps between all parts of configuration space, this acceptance criterion is only of subordinate importance. As we will see below this is still the case for the small cluster sizes studied here, and we therefore simply accept all generated cluster structures within a predefined energy range of interest above the ground-state isomer.

\subsection{Sampling efficiency}

The intended performance analysis requires a well-defined measure for the success of a sampling run. A common choice for this in the literature is the number of trial moves until the global minimum has been found for the first time. Here, we adapt this criterion to the stated goal of identifying not only the global minimum, but also all relevant energetically lowest-lying isomers. Correspondingly, the considered indicator of sampling efficiency which we aim to optimize is the number of moves $N$ until all relevant isomers have been found at least once, where of course one needs to define what a relevant isomer is ({\em vide infra}). While certainly a useful measure for the performance of the employed BH moves, it should still be stressed that due to the slightly varying number of geometry steps for the local relaxation of each trial structure, $N$ is only roughly proportional to the total computational cost of the first-principles BH run. 

Due to the stochastic nature of the BH method, both with respect to the generation of the initial starting structure and the generation of trial structures, $N$ is only a statistically meaningful quantity after averaging over sufficiently many runs. Even for the small cluster sizes considered here, this implies having to run of the order of 100 different first-principles BH runs to obtain a $N_{\rm av}$ that is converged to within $\pm 1$, and this for each BH setting (e.g. move distance or distribution) one wants to analyze. Since this straightforward approach quickly becomes computationally involved, we instead resort to the concept of a ``hopping matrix'' $h$, which summarizes the transition probabilities between all isomers under the chosen BH settings. Specifically, the matrix element $h_{ij}$ is then the normalized probability to jump from the local minimum $i$ to the local minimum $j$. If all local minima are explicitly accounted for, one obviously has the condition 
\begin{equation}
\sum_j h_{ij} = 1 \quad .
\label{eq3}
\end{equation}
Assuming that the matrix $h_{ij}$ is completely known, a sufficiently large number of sampling runs starting in random isomers can be readily simulated entirely on the basis of these transition probabilities without the need for further first-principles calculations, let alone that the individual matrix elements aka transition probabilities provide valuable insight into the sampling process and efficiency. Notwithstanding, with a rapidly growing number of isomers with cluster size this approach merely shifts the computational burden of an increasing number of direct BH runs to the equally expensive computation of an exploding number of hopping matrix elements, i.e. converged transition probabilities. Yet, below we will show that an approximate, but for our purposes sufficient determination of $N_{\rm av}$ is possible by restricting the explicit calculations to a limited number of hopping matrix elements. 

\begin{figure}
\centering
\includegraphics[width=3.7cm,angle=-90]{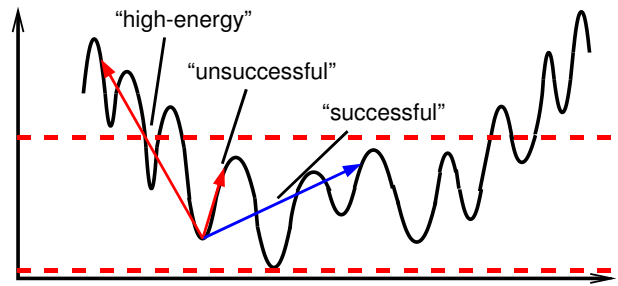}
\caption{(Color online) Schematic illustration of successful, unsuccessful and high-energy trial moves in the BH scheme. The horizontal dashed (red) lines indicate the targeted energy window entering the acceptance criterion.}
\label{fig2}
\end{figure}

In order to further analyze the obtained performance data, it is useful to disentangle the different possible outcomes of a trial move. First of all, the system might relax back into the structure from which the trial move has been performed so that in terms of isomer information nothing has been gained. Correspondingly, we denote such a move as unsuccessful, cf. Fig. \ref{fig2}, and define the fraction of hitherto unsuccessful moves $\alpha_{\rm unsucc.}$ as
\begin{equation}
  \alpha_{\rm unsucc.} \;=\; \frac{N_{\rm unsucc.}}{N} \quad ,
  \label{unsucc}
\end{equation}
where $N_{\rm unsucc.} < N$ is the number of unsuccessful moves during the run. Even if the trial move leads to a different local minimum, the move might still be rejected due to the acceptance criterion, if the new minimum is higher up in energy. The fraction of moves rejected on this basis is defined as
\begin{equation}
  \alpha_{{\rm high}E} \;=\; \frac{N_{{\rm high}E}}{N} \quad ,
  \label{highE}
\end{equation}
where $N_{{\rm high}E} < N$ is the corresponding number of rejected moves. Only the remaining fraction
\begin{equation}
  \alpha_{\rm succ.} \;=\; 1 - \alpha_{\rm unsucc.} - \alpha_{{\rm high}E} \quad
  \label{succ}
\end{equation}
are successful moves at least in the sense that they bring the algorithm to a different minimum out of which the next trial move is performed, albeit not necessarily leading to a minimum that had hitherto not yet been sampled. Just as in the case of $N_{\rm av}$, it only makes sense to analyze the fractions $\alpha_{\rm unsucc.,av}$, $\alpha_{{\rm high}E,{\rm av}}$ and $\alpha_{\rm succ.,av}$ once averaged over sufficiently many different BH runs.

\section{Performance analysis for small cluster sizes}

Our performance analysis concentrates on small clusters formed of Si and small clusters formed of Cu atoms. Both systems have already been subject to extensive theoretical studies and are therefore natural choices for the intended benchmarking. Extensive work on small silicon clusters has both been carried out using wavefunction-based techniques \cite{raghavachari88, zhu03} and DFT \cite{yoo05,hellmann07}. Databases for small silicon isomers can e.g. be found in Refs. \onlinecite{cambridge_database,hellmann07b}. Recent works on small copper clusters using {\em ab initio} methods are e.g. Refs. \onlinecite{yang06,massobrio98,calaminici00,jug02,yang05}. The choice of these two materials is further motivated by their different chemistry, which can be characterized as more covalent and directional in the case of Si, and more metallic in the case of Cu. We therefore expect the direct comparison of results obtained for Si$_7$ and Cu$_7$ to reflect a possible material-specificity of the findings, while an additional comparison of the results obtained for Si$_7$ and Si$_{10}$ aims at assessing the variation with cluster size in the range where due to the limited dimensionality of the configuration space the BH acceptance criterion does not play much of a role ({\em vide infra}).

\subsection{Existence of dominant isomers}

\begin{figure*}[ht]
  \centering
  \subfigure
  {
    \includegraphics[width=4cm, angle=-90]{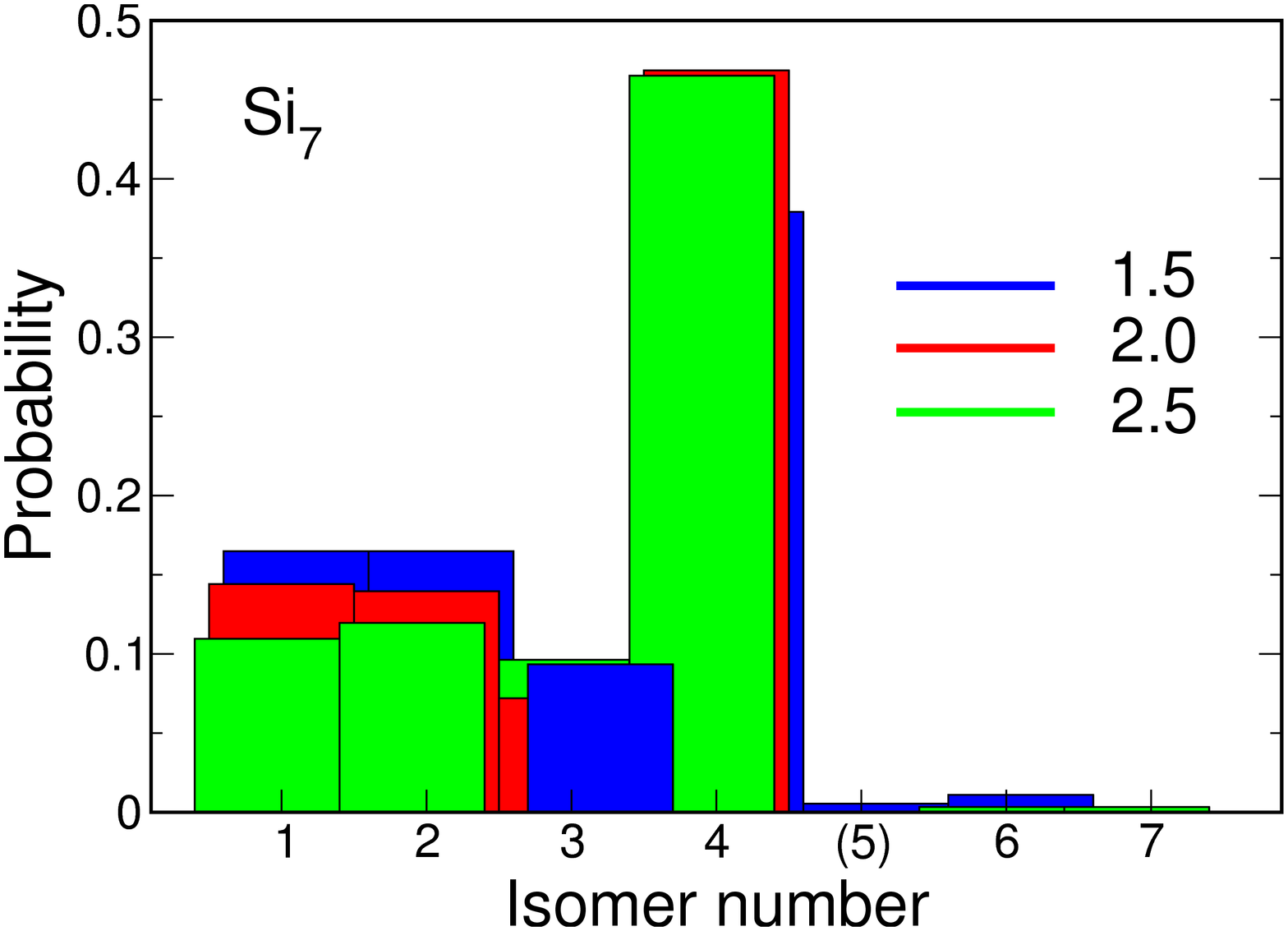}
  }
  \subfigure
  {
    \includegraphics[width=4cm, angle=-90]{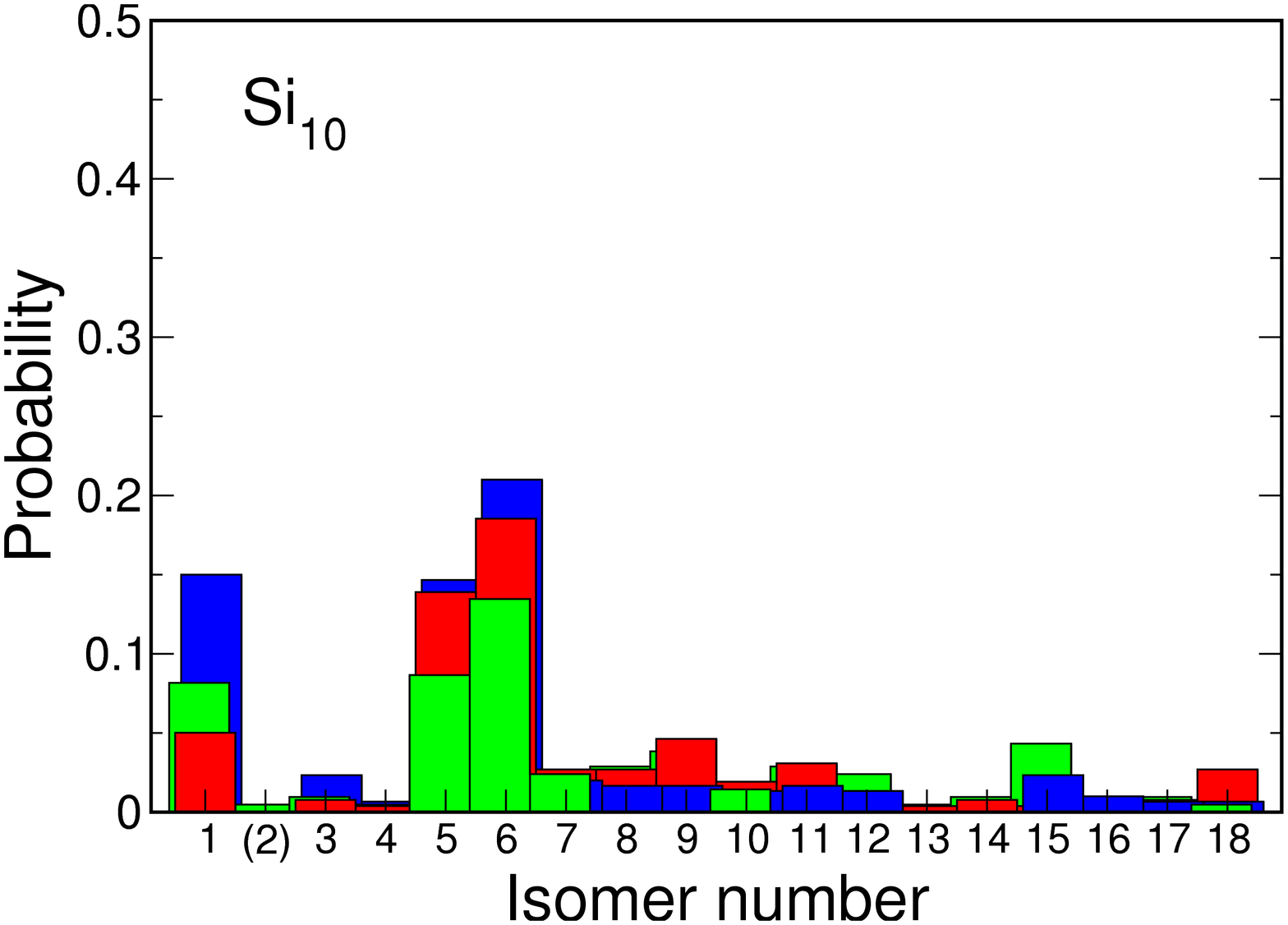}
  }
  \subfigure
  {
    \includegraphics[width=4cm, angle=-90]{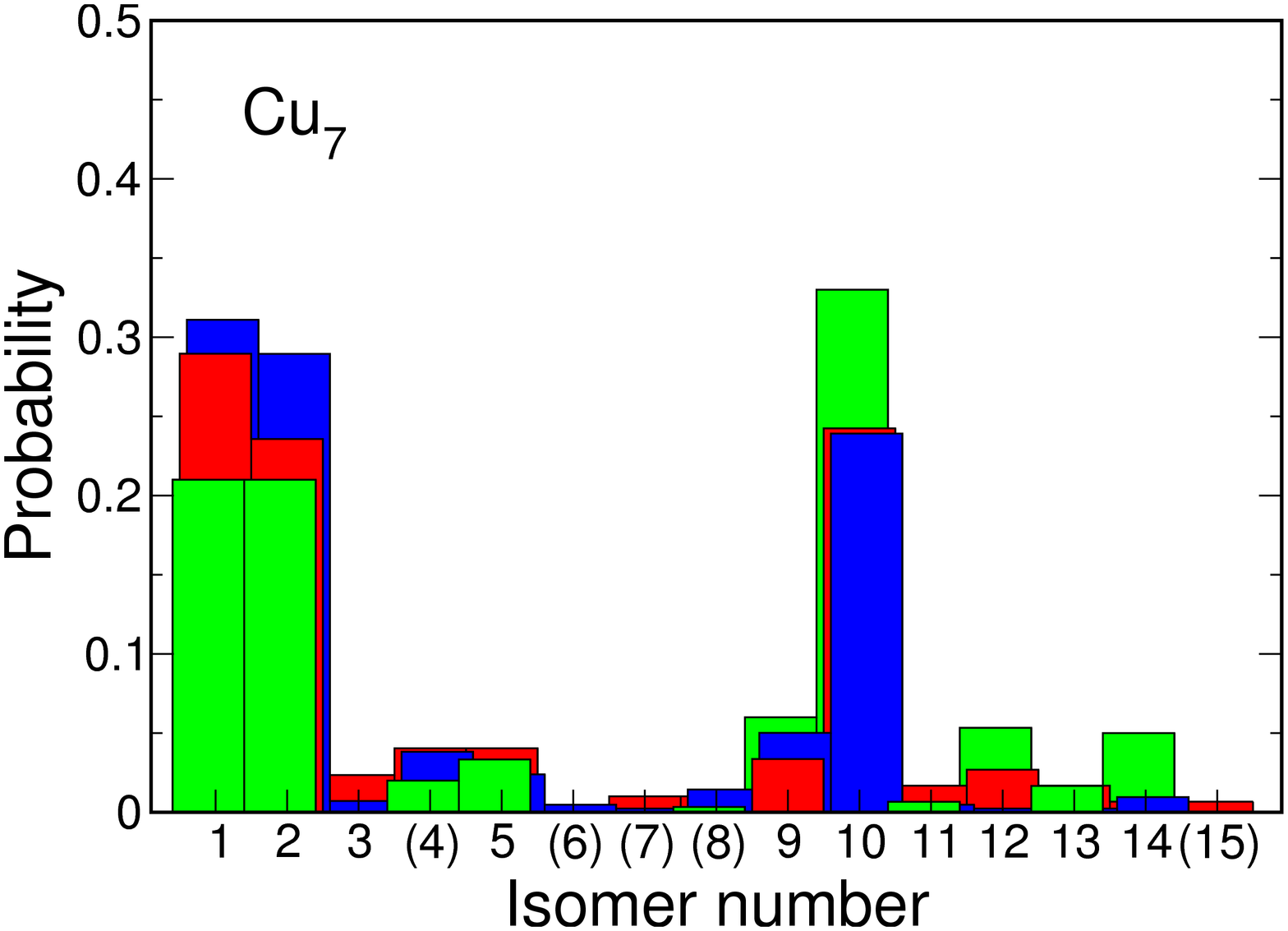}
  }
  \caption{Histograms of the probability with which trial moves end up in the
  lowest-energy isomers of Si$_7$, Si$_{10}$ and Cu$_7$. The identified
  isomers are numbered with decreasing stability, with isomer \#1
  corresponding to the identified ground-state and those isomers shown with bracketed
  numbers revealed as unstable by an {\em a posteriori}
  vibrational analysis (see text). The histograms comprise all isomers found
  in an energy range up to 2\,eV above the ground-state isomer, as obtained
  from long BH hopping runs using single-particle moves and normally distributed move
  distances around the average values $r_{\rm o} = 1.5, 2.0$\, and $2.5$. The 
  geometric structures behind the truly
  stable isomers in an energy range up to 1\,eV above the identified
  ground-state are summarized in Figs. \ref{fig5}-\ref{fig7}.
 } 
  \label{fig3}
\end{figure*}

\begin{figure}[ht]
  \centering
  \includegraphics[width=3.5cm, angle=-90]{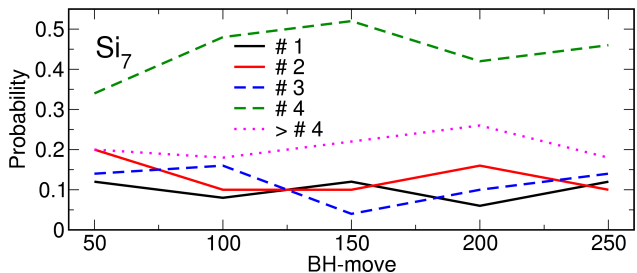}
  \caption{Probabilities for the lowest-energy isomers of $\rm Si_7$
  as in Fig. \ref{fig3}. Shown is the evolution when binning the histogram
  entries over consecutive sampling periods containing 50 moves each using single-particle moves and normally distributed move
  distances around the average values $r_{\rm o} = 2.5$. Entries
  for all isomers higher in energy than isomer \#4 are bundled into one entry
  labeled ``$>$\,\#4''.}
  \label{fig4}
\end{figure}

As a prelude to the actual performance analysis we present in Fig. \ref{fig3} the histograms of the number of times with which low-energy isomers were identified in long BH runs for the three systems addressed, i.e. Si$_7$, Si$_{10}$, and Cu$_7$. Each run consisted of several hundred unconditionally accepted moves and was carried out until the shape of the histogram, i.e. the normalized probability with which the different low-energy isomers are identified, was fully converged. In all cases the evolution towards convergence was rather uniform as demonstrated by Fig. \ref{fig4} for Si$_7$, which presents the histogram entries binned over consecutive sampling periods containing 50 moves each. Apparently, the ratios of the histogram entries for each sampling period are roughly the same. In view of the overall still limited system dimensionality and concomitant small number of low-energy isomers, a natural interpretation for this is that the employed moves enable jumps between any parts of the PES. In this situation, a simple acceptance criterion that unconditionally accepts moves within a pre-defined energy range and rejects all others is then sufficient to separately assess the dependence of the algorithm efficiency on the move parameters.

Even though Fig. \ref{fig3} comprises the data obtained using single-particle moves with three quite different move
distances it is interesting to observe that some isomers are always sampled much more often than others. For Si$_7$ for example, more than one third of all executed moves in the BH runs ended up in the isomer structure labeled \#4, regardless of the actual move distance employed. In the case of collective moves, the corresponding histograms look qualitatively the same so that the existence of such ``preferred'' isomers, which we will henceforth term dominant isomers, seems even independent of the specific move class employed. In this respect, one should mention that some of the isomers listed in Fig. \ref{fig3} turned out to be unstable in the concluding vibrational analysis and are correspondingly not further considered below. Distinguishing and discarding these structures, which correspond either to flat or saddle-point PES regions, directly in the BH run is unfortunately impossible as it would imply a prohibitive computational cost when performing a vibrational analysis immediately after each trial move. As apparent from Fig. \ref{fig3} the total number of times in which the BH runs end up in such unstable structures is at least not too large, so that the actual computational time wasted is small. The one notable exception is isomer \#4 of $\rm Cu_7$, which exhibits small imaginary eigenmodes, but is sampled about as frequently as the truly stable isomer \#5. Since the algorithm thus spends some appreciable time in this basin, we retained isomer \#4 in the ensuing performance analysis despite its instability.

\begin{figure}
  \includegraphics[width=7cm, angle=-90]{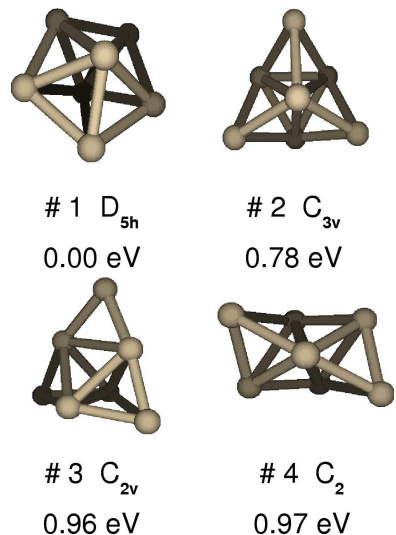}
  \caption{Identified stable $\rm Si_7$-isomers in the energy range up to 1\,eV above the ground-state. The isomer numbering follows the one of Fig. \ref{fig3} and reflects the decreasing cluster stability as indicated by the stated energies relative to the ground-state isomer \#1.}
  \label{fig5}
  \centering
\end{figure}

\begin{figure}
  \includegraphics[width=7cm, angle=-90]{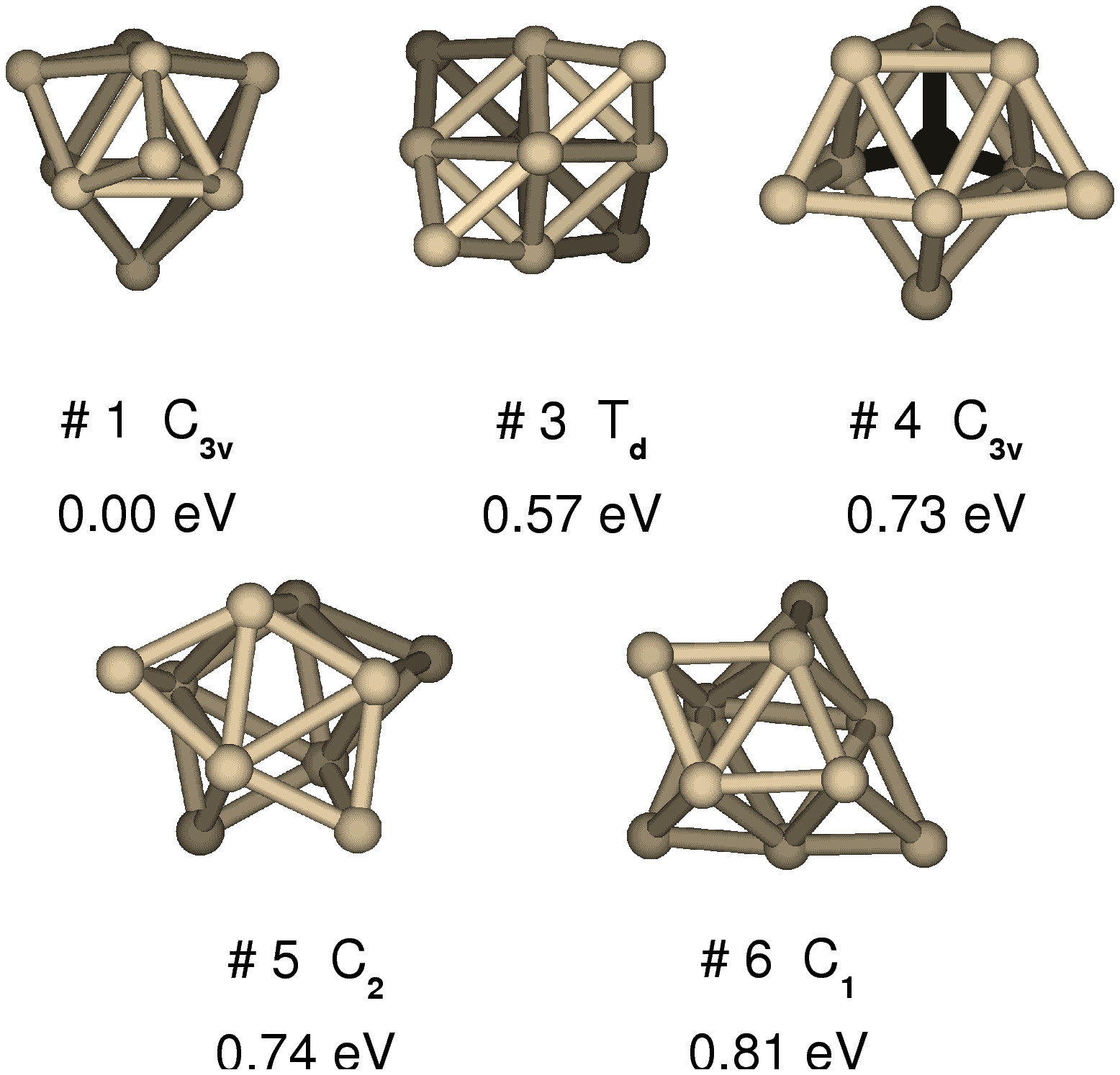}
  \caption{Identified stable $\rm Si_{10}$-isomers in the energy range up to 1\,eV above the ground-state. The isomer numbering follows the one of Fig. \ref{fig3} and reflects the decreasing cluster stability as indicated by the stated energies relative to the ground-state isomer \#1.}
  \label{fig6}
  \centering
\end{figure}

\begin{figure}
  \includegraphics[width=7cm]{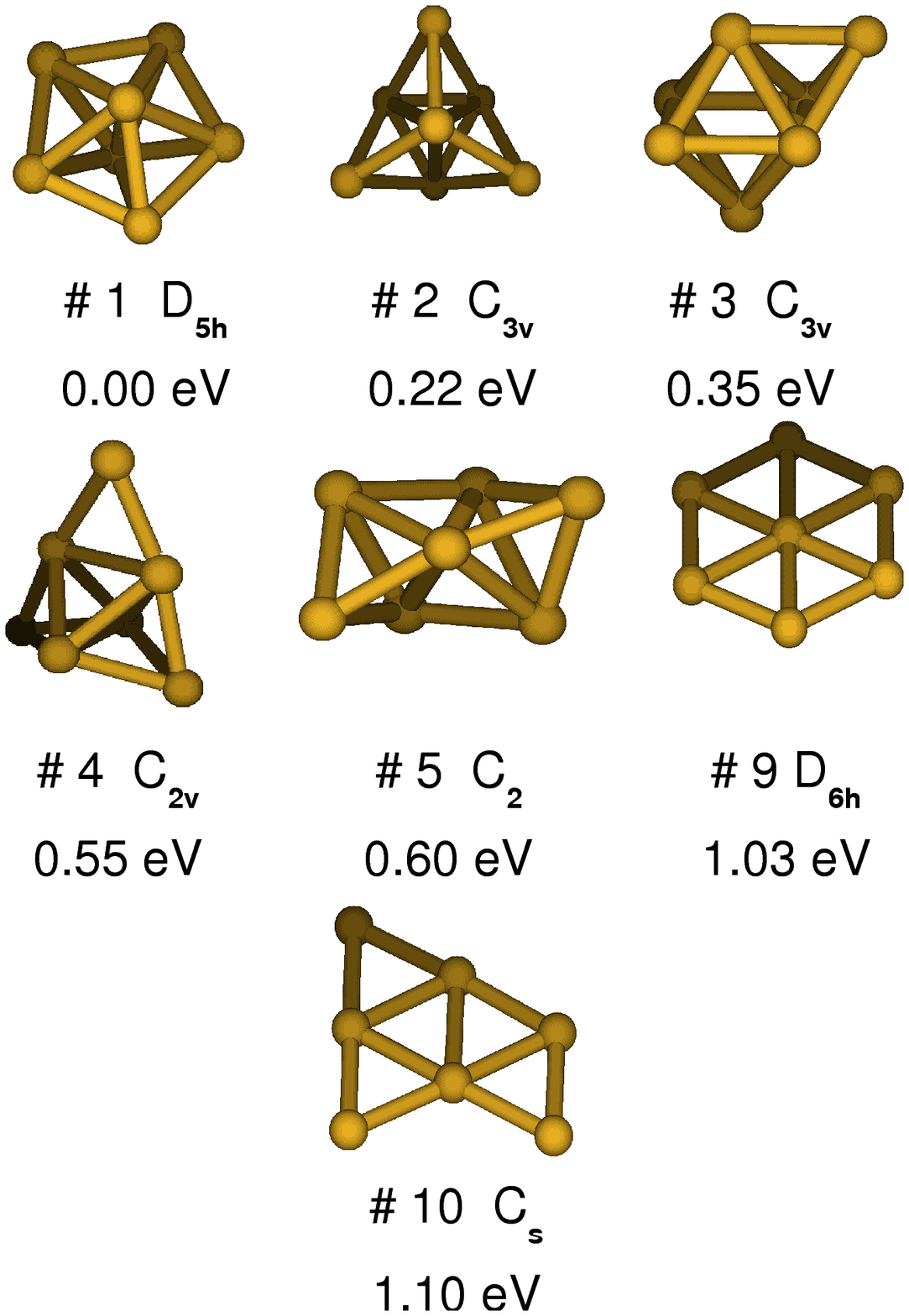}
  \caption{ Identified stable $\rm Cu_7$-isomers in the energy range up to 1.1\,eV above the ground-state. The isomer numbering follows the one of Fig. \ref{fig3} and reflects the decreasing cluster stability as indicated by the stated energies relative to the ground-state isomer \#1. Note that isomer \#4 exhibits small imaginary eigenmodes but is nevertheless retained in the performance analysis, see text.}
  \label{fig7}
  \centering
\end{figure}

One immediate rationalization for the existence of dominant isomers is that their corresponding basin of attraction on the PES is huge and thus hit by the trial moves many times. Inspection of the geometric structures of the lowest-energy isomers for the three systems as summarized in Figs. \ref{fig5} - \ref{fig7} points, however, at a second potential reason. Many of the dominant
isomers correspond to rather low-symmetry structures, e.g. isomer \#4 for Si$_7$, isomer \#6 for Si$_{10}$ or isomer \#10 for Cu$_7$. In terms of the PES, these low-symmetry
structures possess a larger number of local minima than the symmetric ones \cite{wales00}, and it is this multiplicity, and not necessarily only the size of the basin of attraction of each individual minimum that is responsible for the large number of times with which the BH algorithm yields the corresponding isomer. This relation to the underlying PES shape also motivates why certain isomers are dominant irrespective of the employed move class. Any general
purpose move class that enables unbiased jumps to anywhere on the PES should be similarly affected by a varying size or multiplicity of the different basins of attraction. This is an important point as an at first glance appealing approach to improve the efficiency of BH sampling would be to reduce the number of times that the algorithm gets stuck in always the same dominant isomers and instead aim to increase jumps into the rare minima. Within the understanding of the relation to the PES topology it seems unlikely that this can be realized without either resorting to moves that are specifically tailored to the system at hand or make use of local PES information. At least for the limited isomer number of the small cluster sizes studied here, the main bottleneck of purely stochastic moves is thus that the algorithm will often revisit the same dominant isomers. In this situation, the overall performance is then dictated by the way it can deal with these dominant isomers, e.g. how efficiently it can hop out of them.

\subsection{Approximate hopping matrix}

On the basis of the histograms presented in Fig. \ref{fig3} we can now specify which of the energetically lowest-lying isomers are the target of the sampling runs. In the general case, this would be dictated by the physics of the problem at hand, e.g. prescribing that the sampling should yield the ground-state isomer, as well as all isomers in a certain energy range above
it. In view of the discussion above, it is clear that the overall sampling performance will in any case be governed by the dominant isomers involved, since the algorithm spends most of its time jumping out of these minima. For the intended performance analysis we therefore choose as the sampling target the identification of all dominant isomers determined in the histogram BH-runs. As indicator of the sampling efficiency we correspondingly focus on the number of moves $N$ until all of these dominant isomers are found at least once. In the case of Si$_7$ and Si$_{10}$ the dominant isomers are included in an energy range up to 1\,eV above the ground-state as apparent from Figs. \ref{fig3}, \ref{fig5} and \ref{fig6}. In the case of Cu$_7$, this energy range is slightly extended to 1.1\,eV above the ground-state to also include the dominant isomer \#10, cf. Figs. \ref{fig3} and \ref{fig7}. 

With a thus defined sampling target the BH acceptance criterion employed is to unconditionally accept trial moves that lead into any isomer in the corresponding energy window, and to unconditionally reject any trial move that leads into an isomer that is higher in energy. It would only be necessary to change the latter to some, e.g. Boltzmann weighted, conditional acceptance rule, if a multiple-funnel type PES would necessitate passages via such higher-energy isomers. However, as discussed above this is not the case for the systems studied here. In terms of the hopping matrix corresponding energy-window BH runs require only the knowledge of a limited number of hopping matrix elements. Definitely required are the transition probabilities between any of the targeted low-energy isomers. Since trial moves into higher energy isomers are rejected, it suffices in addition to know the overall probability to jump from each one of the low-energy isomers into any of the higher energy ones, without the need to further resolve the latter. For the example of Si$_7$ the targeted energy window comprises four different isomers, and energy-window BH runs can therefore be simulated on the basis of 20 hopping matrix elements: 16 transition probabilities between any of the four different low-energy isomers, as well as one hopping matrix element per low-energy isomer that describes the sub-summed transition probability to jump out of the isomer into any of the higher energy ones. 

For a specified BH setting (i.e. fixed move type and fixed technical move parameters) we obtain the required hopping matrix elements by performing a fixed number of trial moves out of each of the low-energy isomers, recording the probabilities with which the moves led into each of the other low-energy isomers or any of the higher-energy ones. After 100 moves these probabilities are converged to within $\pm 0.1$ at a confidence interval at the level of 95\%, which we found to be sufficient for the conclusions put forward below. With the thus determined hopping matrix, a large number of energy-window BH runs from different starting isomers and with different random number sequences can be quickly simulated without the need for further first-principles calculations. This allows to arrive at a properly averaged number $N_{\rm av}$ of moves required to determine all low-energy isomers at least once, albeit with the disadvantage that the transition probabilities are only known within the confidence interval of $\pm 0.1$. To account for the latter, we therefore randomly varied the individual hopping matrix elements within this uncertainty range and under the constraint of Eq. (\ref{eq3}). Determining the $N_{\rm av}$ for several thousands of correspondingly created hopping matrices, we finally quote below the average value together with error bars given by the standard deviation.  

This remaining uncertainty incurred from the approximate hopping matrix procedure does not affect any of the trend conclusions made below, yet on the other hand leads to quite some reduction in the computational effort: In order to determine a converged $N_{\rm av}$ for the systems studied here typically required an averaging over some hundred BH runs starting from different initial isomers and with different random number sequences. As shown below in the range of settings studied $N_{\rm av}$ is of the order of 10-40, so that a straightforward determination of $N_{\rm av}$ by averaging over individual first-principles BH runs would require a few thousand trial moves, with a corresponding number of first-principles energy and force evaluations. For the described hopping matrix based approach, however, only 100 moves out of each of the few low-energy isomers need to be done on the basis of first-principles calculations. Since the ensuing hopping matrix based simulations are computationally undemanding, this significantly reduces the overall computational cost and provides furthermore detailed data on the sampling process in form of the individual hopping matrix elements.

\subsection{Dependence on move parameters}

\begin{figure*}
  \centering
  \subfigure
  {
    \includegraphics[width=7cm, angle=-90]{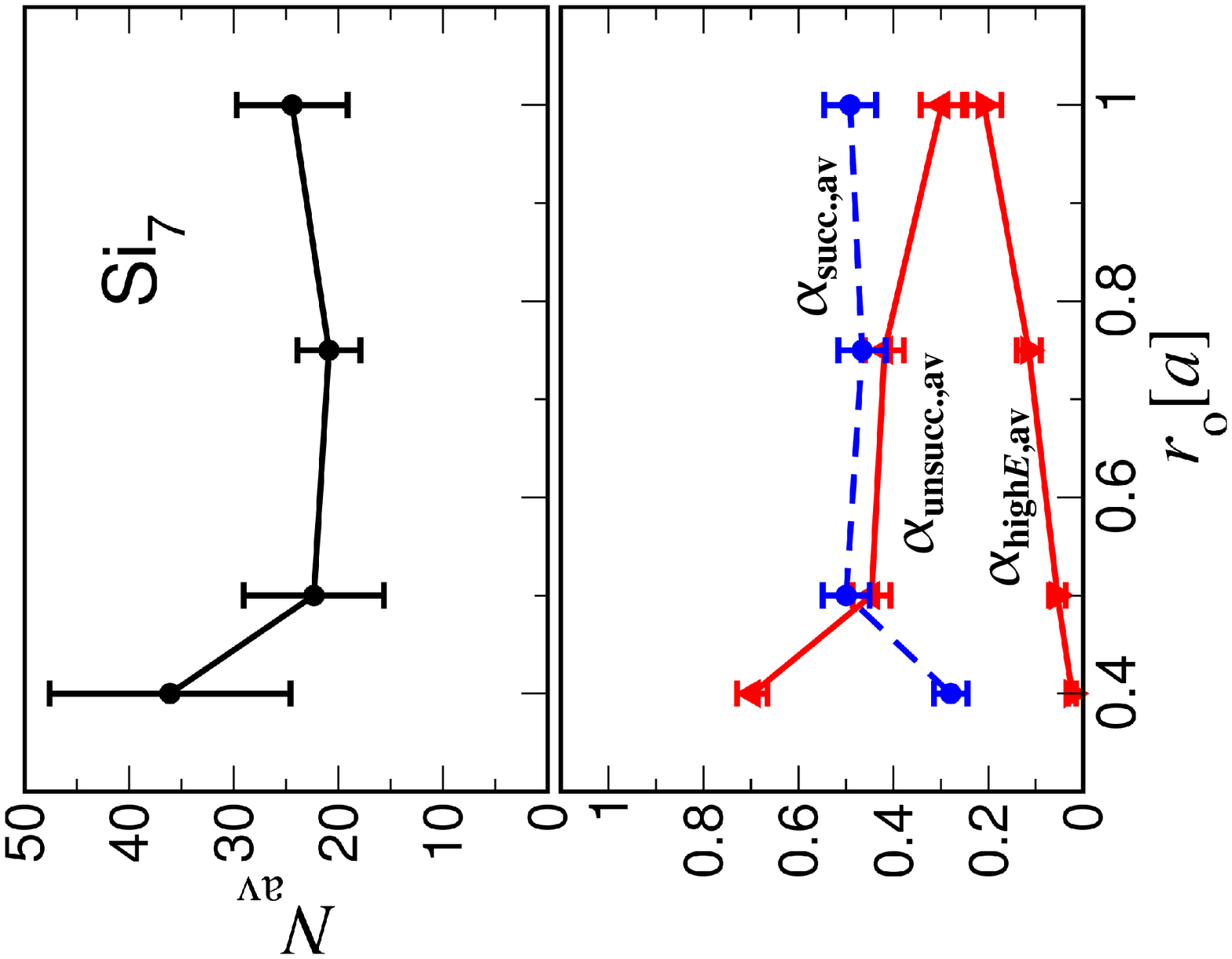}
  }
  \subfigure
  {
    \includegraphics[width=7cm, angle=-90]{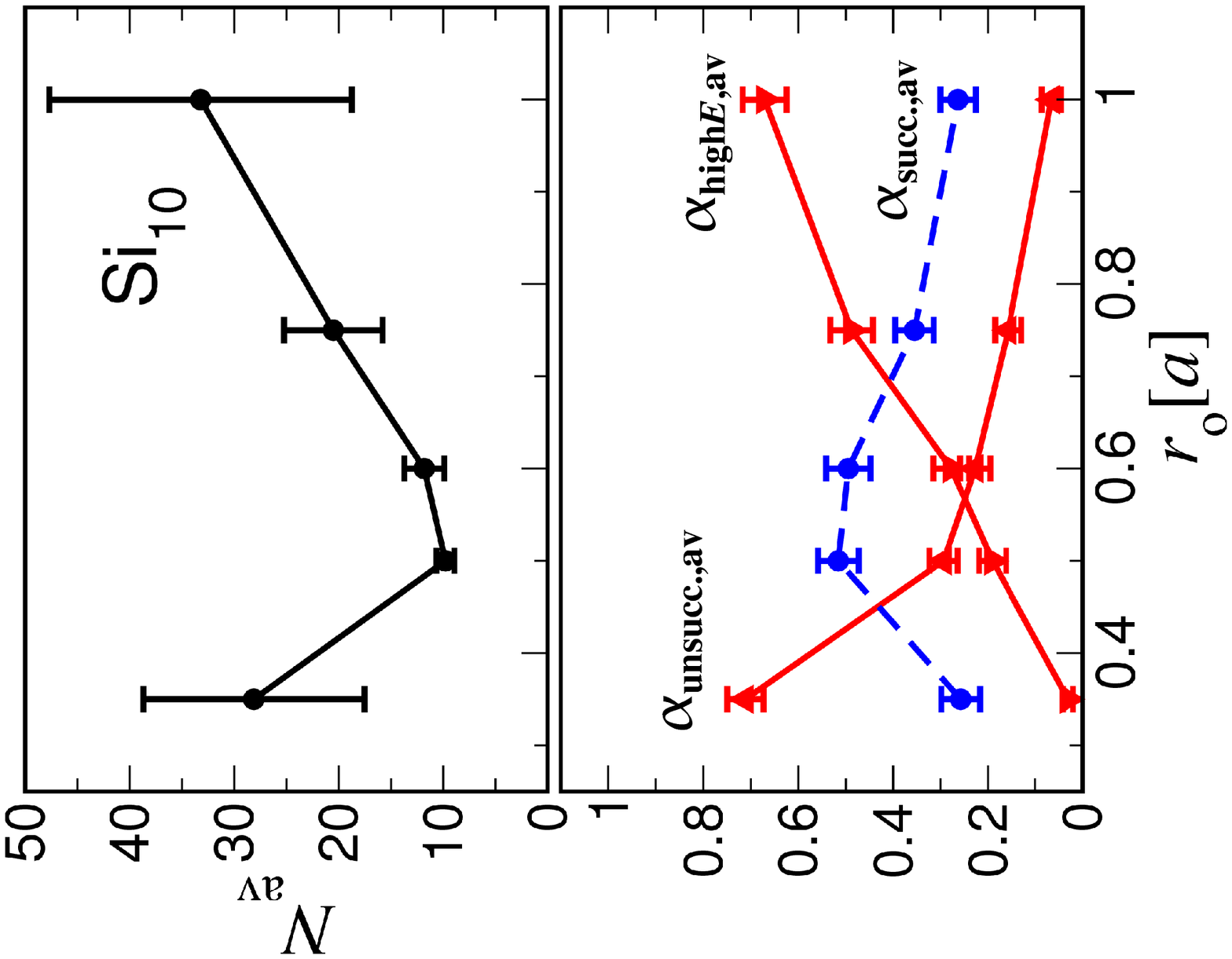}
  }
  \subfigure
  {
    \includegraphics[width=7cm, angle=-90]{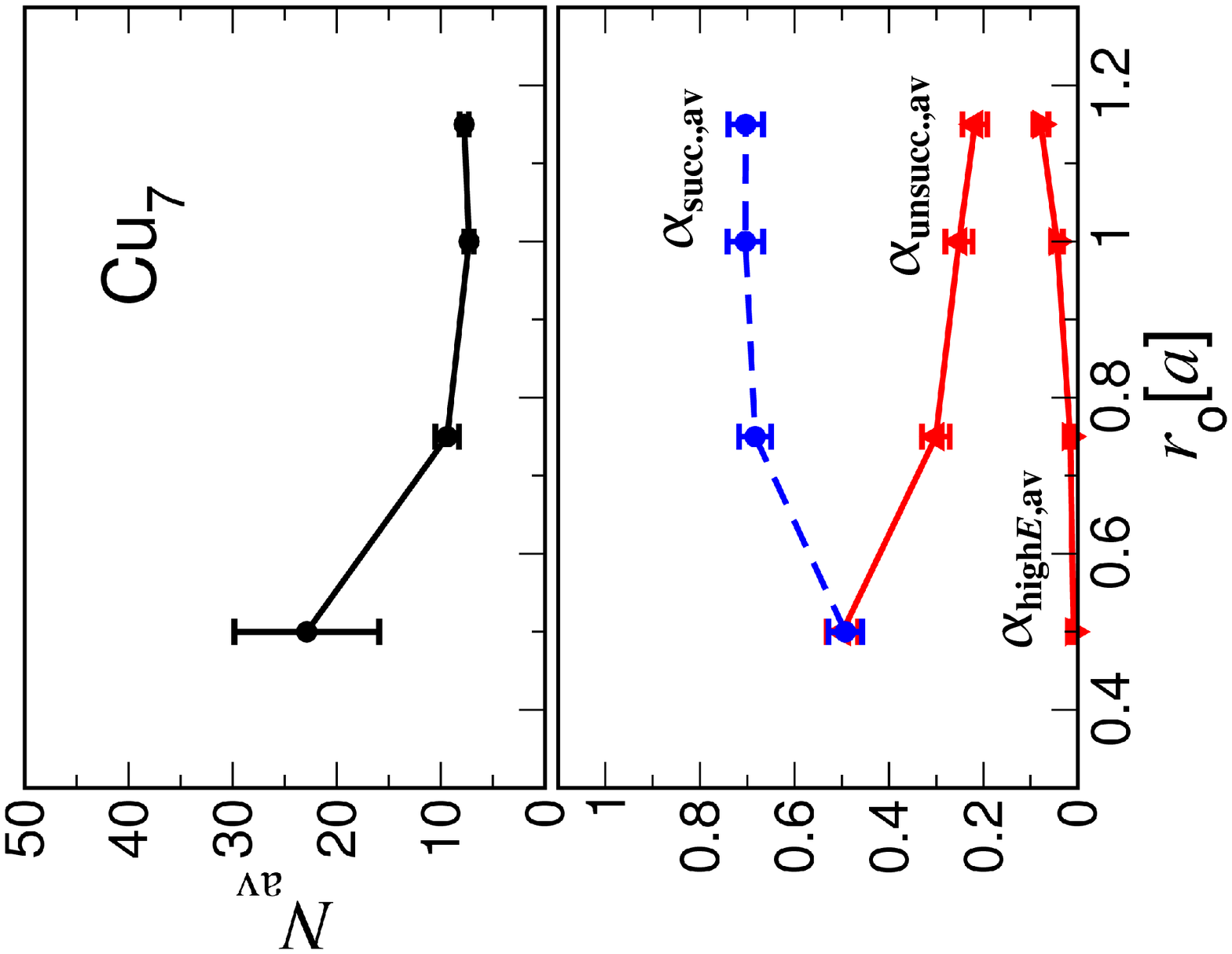}
  }
  \caption{(Color online) Performance analysis of BH runs for Si$_7$, Si$_{10}$, and Cu$_7$, using collective moves and a 
  normal distribution for the atomic displacements. Upper panel: Variation of the average number of moves $N_{\rm av}$ 
  required to determine the low-energy dominant isomers with the average move distance $r_{\rm o}$ (see text). Lower
  panel: Corresponding variation of the fraction of unsuccessful moves $\alpha_{\rm unsucc.,av}$, of moves into high energy 
  isomers $\alpha_{{\rm high}E,{\rm av}}$, and of successful moves $\alpha_{\rm succ.,av}$, cf. Eqs. (\ref{unsucc}-\ref{succ}). 
  The error bars reflect the uncertainty due to the employed approximate hopping matrix procedure (see text).}
  \label{fig8}
\end{figure*}

We begin the analysis with the performance data obtained for collective moves and a normal distribution for the atomic displacements. Figure \ref{fig8} compiles the corresponding results and reveals a similar dependence of $N_{\rm av}$ on the average move distance for the three systems. In all cases, a too small value of $r_{\rm o}$ leads to a large move number required to determine the low-energy isomers. With increasing $r_{\rm o}$ the performance gets better, and goes through an optimum that is more pronounced for Si$_{10}$ than for the two smaller systems. This overall dependence is well rationalized by analysing the move fractions defined in Eqs. (\ref{unsucc}-\ref{succ}) above. Not surprisingly, the bad performance at too small move distances results from the inability of the algorithm to escape from the present basin of attraction, as reflected by a fraction $\alpha_{\rm unsucc.,av}$ approaching unity, cf. Fig. \ref{fig8}. With increasing move distances, this fraction of unsuccessful moves decreases and the overall performance improves. Interestingly, within the studied range of move distances $\alpha_{\rm unsucc.,av}$ only quickly decays to around zero for Si$_{10}$, whereas for the two smaller systems it seems to level off at a finite value. This behavior arises from the afore discussed multiplicity of some of the dominant isomers. In terms of the hopping matrix, $\alpha_{\rm unsucc.,av}$ is just the average of the diagonal elements $h_{ii}$ for the different isomers $i$ weighted by the corresponding histogram entries, where $h_{ii}$ gives the probability that a hop out of isomer $i$ has unsuccessfully relaxed back into it. Inspecting these diagonal elements for the different isomers separately we find only the elements of the most symmetric isomers to vanish with increasing move distance. On the contrary, for the least symmetric isomers the corresponding hopping matrix elements stay almost constant over the range of move distances studied. The rational is that by choosing a sufficiently large move distance, the system can be prevented from relaxing back into the previous PES minimum, but not from jumping into another symmetry-equivalent basin of attraction. The value at which $\alpha_{\rm unsucc.,av}$ saturates is therefore system-dependent and governed by the symmetry properties of the dominant isomers in the targeted energy range.

This finite energy range of interest, and the correspondingly applied acceptance criterion, introduces a second ruling factor for the overall efficiency of the algorithm. As apparent from Fig. \ref{fig8}, the fraction of rejected moves that has led to isomers outside the targeted energy window rises monotonically with increasing move distance. Naively equating the move distance with the perturbation induced by the trial move, this is somehow intuitive. In view of the rapidly increasing total number of isomers with system size one may further consider the steeper increase of $\alpha_{{\rm high}E,{\rm av}}$ for Si$_{10}$ as reflecting the increasing fraction of isomers that fall outside the defined low-energy window in this larger system. Even when for instance only focusing on the energy range up to 2\,eV above the identified ground-state isomer, the long BH runs behind the histograms shown in Fig. \ref{fig3} found only 2 and 4 stable isomers outside the presently targeted low-energy window for Si$_7$ and Cu$_7$, respectively, but already 12 in the case of Si$_{10}$. While the fraction of unsuccessful moves is thus the bottleneck at short move distances, so is the fraction of moves outside the energy window at large distances, and this will become more severe with increasing system size or when reducing the targeted energy range.

The variation of the fraction of successful moves $\alpha_{\rm succ.,av}$ with move distance is determined by the opposing trends of $\alpha_{\rm unsucc.,av}$ and $\alpha_{{\rm high}E,{\rm av}}$, cf. Eq. (\ref{succ}), and exhibits a clear correlation with the obtained performance. As obvious from Fig. \ref{fig8}, the average number of moves $N_{\rm av}$ required to find all low-energy isomers is least when the fraction of successful moves is maximized. This is the case when the move distance is large enough to efficiently lead the system out of the present basin of attraction, but not too large to yield a high energy isomer outside the targeted energy window. With the much more pronounced increase of $\alpha_{{\rm high}E,{\rm av}}$ for Si$_{10}$ this gives rise to a narrowly defined range of optimum move distances, which is concomitantly also shifted to smaller values compared to the two smaller systems. As apparent from the error bars in Fig. \ref{fig8} this overall performance behavior and its analysis in terms of the different move fractions $\alpha_{\rm unsucc.,av}$, $\alpha_{{\rm high}E,{\rm av}}$, and $\alpha_{\rm succ.,av}$ is robust against the uncertainty introduced by the approximate hopping matrix procedure. It is furthermore equivalently obtained for the other move schemes investigated, i.e. single-particle vs.\,collective moves involving atomic displacements following a uniform or normal distribution around the average distance $r_{\rm o}$. 

\begin{table}
\caption{\label{table1}
Lowest obtained average number of moves $N_{\rm av}$ to identify the targeted low-energy isomers of Si$_7$, Cu$_7$, and Si$_{10}$ using different trial move schemes. Quoted are the values and the corresponding average move distance $r_{\rm o}$, that within the finite resolution computed comes closest to the optimum settings. Within the understanding gained from the two smaller systems, the run for Si$_{10}$ using single-particle moves with a uniform distribution was not performed.}
\begin{ruledtabular}
\begin{tabular}{ll|cc|cc}
                &        & \multicolumn{2}{c}{normal distribution} & \multicolumn{2}{c}{uniform distribution} \\
                &        & $r_{\rm o}$    & $N_{\rm av}$           & $r_{\rm o}$    & $N_{\rm av}$            \\[0.1ex] \hline
single-particle & Si$_7$ &  1.5           &  20                    &  1.5           & 31                      \\
                & Cu$_7$ &  2.0           &  9                     &  1.5           & 20                      \\
             & Si$_{10}$ &  1.5           &  10                    &   $-$          & $-$                     \\[0.5ex]
collective      & Si$_7$ &  0.75          &  21                    &  0.75          & 18                      \\
                & Cu$_7$ &  0.75          &  9                     &  0.75          & 8                       \\
             & Si$_{10}$ &  0.5           &  10                    &  0.5           & 15                      \\[0.5ex]
\end{tabular}
\end{ruledtabular}
\end{table}

Table \ref{table1} summarizes for the different schemes the obtained lowest values for $N_{\rm av}$ at the move distance that within the finite resolution computed comes closest to the optimum setting. Starting with single-particle moves we observe a significantly better performance for displacements that are drawn from a normal distribution peaked around the average value $r_{\rm o}$. This demonstrates that for the systems studied the wide range of move distances featured by the uniform distribution is not advantageous for the sampling. Instead, there is indeed an optimum atomic displacement on which the employed moves should focus. This is consistent with the understanding of the limiting factors at too small and too large displacements developed above, and in this respect we believe this result to be more generally valid. Our interpretation for the much less pronounced performance difference between uniform and normal distribution in case of collective moves, cf. Table \ref{table1}, is correspondingly that even when displacing all atoms by random distances that are uniformly distributed over a wide range there is a certain probability that at least one of these distances comes close to the optimum value. Regardless of the other displacements, for the small systems studied this one near-optimum displacement is then sufficient for an efficient sampling as also indicated by the essentially identical performance of single-particle and collective moves obeying a normal distribution. This said, we nevertheless note that another factor entering here is that the optimum $r_{\rm o}$ in case of collective moves is much shorter, with a concomitant reduction in the width of the employed uniform distribution and therewith of the difference between the two distributions studied. 

The shorter values for the optimum displacement in case of collective moves are intuitive considering that in order to change the geometric configuration significantly the more atoms are involved, the less each atomic position needs to be disturbed. It is, however, intriguing to see that in terms of the dimensionless quantity $r_{\rm o}$ the optimum values obtained for the three investigated systems are rather similar, both in case of single-particle moves and in case of collective moves. In view of the different chemistry of Si and Cu, this suggests that employing the computed dimer bond length $a$ as natural unit for the move distance is useful for these monoatomic systems. While the general philosophy of the present work aims at an optimization of the sampling efficiency, a tentative generalization of our findings would therefore nevertheless be that setting the move distance to somewhere short of the dimer bond length in case of collective moves or at around 1.5 times the dimer bond length in case of single-particle moves is not a bad strategy to achieve already quite decent sampling. In this respect, we also note that the performance variation with $r_{\rm o}$ is in all cases similar to the one illustrated for collective moves with normal distribution in Fig. \ref{fig8}: Over the distance range studied, which was $a$ to 2.5$a$ for single-particle and $a/3$ to $a$ for collective moves, the efficiency of the BH scheme is thus quite robust and varies in most cases much less than an order of magnitude. In light of the discussion concerning the fraction of moves $\alpha_{{\rm high}E,{\rm av}}$ that lead to isomers outside the targeted energy window we expect this variation to become much more pronounced for larger systems or a reduced energy range of interest. In this situation optimization of the move settings will be crucial and the observed and intuitive correlation of the overall performance with the fraction of successful moves may then suitably be exploited to analyze and possibly even adapt the settings of an on-going run. However, as illustrated by the data in Fig. \ref{fig8} aiming at an absolute value for the ratio of accepted trial structures, like the empirical factor one half to achieve good sampling of canonic ensemble averages \cite{wales97,frenkel02}, seems not the right approach. Even though in Fig. \ref{fig8} $\alpha_{\rm succ.,av}$ is at optimum move distance indeed about 50\% for Si$_7$ and Si$_{10}$, it is about 70\% in case of Cu$_7$. Aiming at about 50\% in the latter case would instead result in a move distance that is too short ($0.5a$), at a performance that is by a factor 2-3 worse than at optimum settings, cf. Fig. \ref{fig8}. On the contrary we consistently observe for all studied systems, move types, and displacement distributions that the best performance is reached when the ratio of accepted trial structures is largest. This suggests that algorithms aiming to maximize $\alpha_{\rm succ.,av}$ instead of achieving a preset target value are the right way when thinking about adapting move settings on-the-fly.

\section{Conclusions}

In conclusion we have presented a systematic performance analysis of first-principles basin-hopping runs, with the target to identify all low-energy isomers of small atomic clusters within a defined energy range. As representative and widely employed general-purpose move classes we have focused on single-particle and collective moves, in which one or all atoms in the cluster at once are displaced in a random direction by some prescribed move distance, respectively. For the systems Si$_7$, Cu$_7$, and Si$_{10}$ studied, our analysis shows that there is indeed an optimum move distance and that it is not advantageous for the overall sampling to include partly shorter and partly longer moves. The governing factors leading to this optimum move distance are the inability to escape from the basin of attraction of the present configuration at too short distances and the increased probability to end up in high-energy isomers at too large distances. Despite the distinctly different chemistry of Si and Cu, the obtained optimum move distance is similarly roughly 0.75 times the dimer bond length in case of collective moves or at around 1.5 times the dimer bond length in case of single-particle moves. This suggests the dimer bond length as a useful natural unit for these monoatomic systems and as a simple rule-of-thumb that setting the move distance to the mentioned values should enable relatively decent sampling. This is furthermore supported by the observation of only moderate variations of the overall efficiency over quite a range of move distances away from the optimum values.

From our analysis we expect this variation to become more pronounced with increasing system size or when reducing the targeted energy window. With the then increased necessity to optimize the move settings, a possibility to adapt the latter already during an on-going run would be to exploit the confirmed correlation between sampling performance and fraction of accepted trial structures. The latter quantity is an on-the-fly measurable performance indicator, which according to our data devised algorithms adapting the move settings should strive to maximize, rather than aiming for a prescribed target value. However, for larger systems these ideas require further scrutiny. For the small cluster sizes studied here, the sampling problem is still very modest and the employed single-particle or collective moves enable efficient jumps anywhere in configuration space, as also reflected by the essentially identical performance of the two move classes at optimized settings. With increasing system size this is unlikely to hold, and the actual BH acceptance criterion above the targeted energy window will start to play a role to tackle concomitantly developing multiple-funnel potential-energy surfaces. 

While the here investigated size range up to 10 (or slightly more) atoms might not yet be too challenging from a sampling point of view, it is certainly a range that can no longer be reliably covered by resorting to chemical intuition and testing for 
usual-suspect structures. This holds in particular for systems exhibiting strong Jahn-Teller distortions \cite{gehrke08} and when aiming to identify not only the ground-state, but all low-energy isomers. The in this size range furthermore delicate quantum interplay between structural and electronic degrees of freedom dictates an energetics that is based on computationally intense first-principles calculations. In this respect, the observed performance of the BH algorithm employing two simple, general-purpose move classes is reassuring. For all three systems studied, the low-energy isomers in the range up to about 1\,eV above the ground-state are at near-optimum settings identified with a number of trial moves that is perfectly manageable on present-day capacity compute infrastructures. With the still limited number of metastable structures even for the Si$_{10}$ cluster, this algorithmic performance is bound by frequent revisits to a few dominant isomers. Tracing the latter back to the size or multiplicity of the corresponding basins of attraction on the potential-energy surface it seems unlikely that the performance may be significantly improved by other move classes, unless specifically tailoring the latter to the system at hand or making use of local PES information. Nevertheless, when assessing such more specialized move types (also in view of the much more demanding size range beyond ten atoms) the evaluation should be based on a performance analysis protocol as presented in this work.

\section{Acknowledgements}

Funding within the MPG Innovation Initiative ``Multiscale Materials Modeling of Condensed Matter'' is gratefully acknowledged. We are indebted to Dr. Volker Blum and the FHI-aims development team for useful discussions and technical support.

\end{document}